\documentclass[12pt]{report}
\usepackage{amstex,amssymb}

\renewcommand{\theequation}{\arabic{equation}}

\def\ds{\displaystyle}

\title{Topological Charge in Curved Spacetime}
\author{Joseph Saaty \\
The Union Institutute} 

\begin{document}
\bibliographystyle{alpha}

\pagestyle{plain}

\maketitle

\newpage

\setcounter{page}{-0}
\pagenumbering{arabic}
\pagestyle{plain}
\baselinestretch{2}
\baselineskip=22pt
%%%%%%%%%%%%%%%%%%%%%%%
\begin{abstract}

This paper presents the extension from flat spacetime 
into curved spacetime of the area of theoretical investigation 
that has been known as topological gauge field theory.
The extension here presented is based upon a new derivation 
of the expression for topological charge for bosons and fermions
in flat spacetime, a derivation which
has been presented elsewhere [1].
This new approach was developed because 
the established instanton solution could 
not be extended to curved spacetime.  The new approach can be
extended to curved spacetime
by coupling the major equations of relativistic quantum mechanics
to the scalar curvature.  The coupling here presented, and results 
obtained about the quantization of topological charge, 
had not been possible with the earlier established
instanton solution.

\end{abstract}
%js4.tex follows (formerly ch.5)

\def\theequation{\arabic{equation}}

%BEGIN HERE

\noindent
The Klein-Gordon equation in curved spacetime is 
$(\Box^2 - m^2 - \dfrac 16 R) \phi = 0$.

which is $(\Box^2 - m^2) \phi = 0$ (the Klein-Gordon equation in flat
spacetime) coupled to $(-\dfrac 16 R) \phi$ where R is the scalar curvature.
\\

\noindent
Similarly, the Dirac equation in curved spacetime is 
$(\Box^2 - m^2 - \dfrac 14 R) \phi = 0$.

which is $(\Box^2 - m^2) \phi = 0$ (again, the Klein-Gordon equation in flat
spacetime) coupled in this case to $(-\dfrac 14 R) \phi$ where R is the scalar curvature.

(In the above expressions, $\Box^2$ is 
the special relativistic invariant D'Alembertian.) 
\\

\noindent
\noindent
The Dirac equation in flat spacetime in presence of electromagnetic field
is [1]:
\begin{equation}\label{eq.1}
   \left[ - (-\frac 1i \frac{\partial}{\partial t} + e\phi)^2
   + (\frac 1i \nabla + e\vec{A})^2 + m^2 +  f({x},t)\right]
     \Psi (\vec{x},t) = 0.
\end{equation}

From this formula, it was determined that the topological charge for fermions
in flat spacetime cannot be quantized [1].

%section 6.2 (from JS5.tex follows)
\noindent
\\
{\bf To examine the quantization of topological charge in curved
spacetime for bosons and fermions} 
\\

\qquad By operating on the component of the field vector in curved spacetime, 
\cite{3} we get
\begin{align*}
	\nabla_b\phi_c &= \partial_b\phi_c - \Gamma^d_{bc}\phi_c.\\
\nabla_a\nabla_b\phi_c &= \nabla_a\underbrace{(\partial_b\phi_c - 
    \Gamma^d_{bc}\phi_d)}_{\nabla_b\phi_c}, \\
\nabla_a\nabla_b\phi_c &= \partial_a(\partial_b\phi_c - \Gamma^d_{bc}\phi_c)
- \Gamma^e_{ab}(\partial_e\phi_c - \Gamma^d_{ec}\phi_d)
- \Gamma^e_{ac}(\partial_b\phi_e - \Gamma^d_{be}\phi_d)\\
\phi_d &= g_{dk}\phi^k\\
\phi^k &= g^{kc}\phi_c \ \mbox{thus} \\
\phi_d &= g_{dk} g^{kc}\phi_c \ \mbox{will give us} \\
\nabla_a\nabla_b\phi_c &= \partial_a(\partial_b\phi_c 
- \Gamma^d_{bc}g_{dk}g^{kc}\phi_c)
- \Gamma^e_{ab}(\partial_e\phi_c - \Gamma^d_{ec}g_{dk}g^{kc}\phi_c) \\
& - \Gamma^e_{ac}(\partial_b\phi_e - \Gamma^d_{be}g_{dk}g^{kc}\phi_c)\\
 \phi_e &= g_{e\ell}g^{\ell c}\phi_c  \ \ \mbox{implies that}\\
\nabla_a\nabla_b\phi_c &=  \Big[\partial_a(\partial_b 
- \Gamma^d_{bc}g_{dk}g^{kc})
- \Gamma^e_{ab}(\partial_e - \Gamma^d_{ec}g_{dk}g^{kc}) \\
& - \Gamma^e_{ac}(\partial_b g_{e\ell}g^{\ell c}  - \Gamma^d_{be}g_{dk}g^{kc})\Big]
\phi_c.
\end{align*}
$\phi_c$ is a component of a field vector and it appears on both sides
of the above equation.

\newpage
The solution of the Klein-Gordon equation in the absence of electromagnetic
field $\vec{A}$ can be regarded as one component of Dirac's vector field
solution.
\begin{align*}
  \phi_{\mbox{\small{K-G}}} &= \phi = \phi_c \\
g^{ab}\nabla_a\nabla_b\phi &= \underbrace{g^{ab}\partial_a\partial_b\phi}_{\Box^2\phi}
+ \ \mbox{terms in} \ g^{ab}, \ \Gamma, \ \phi\mbox{s}, \ \mbox{and} \ 
\partial\mbox{s}
\end{align*}
$\Box^2$ is the special relativistic invariant D'Alembertian, therefore

\begin{align*}
    g^{ab} \nabla_a\nabla_b\phi & = \Box^2 + \mbox{function of} \ x \\
     & = \Box^2 + f'(x)
\end{align*}
\noindent
for the minimal coupling case, $(g^{ab} \nabla_a\nabla_b + m^2)\phi  = 0$
becomes $(\Box^2 + m^2 + f'(x))\phi = 0$.  In the presence of electromagnetic
field $\Box^2 \rightarrow (\ds\frac 1i \nabla + e\vec{A})^2 -
(-\ds\frac 1i \dfrac{\partial}{\partial t} + e\phi)^2$ and the above
equation becomes equivalent to (equation 1), which is the equation for the
topological charge for fermions in flat spacetime in presence
of electromagentic field.  
Since the equations for both fermions and bosons in curved spacetime 
reduce to the same equation for fermions in flat spacetime, and
since topological charge for fermions in flat spacetime was shown to
be not quantized [1], we can therefore conclude
that for bosons obeying the
minimal coupling, Klein-Gordon's equation, the topological
charge in curved spacetime is not quantized.

Similarly, for the case of bosons obeying the coupling to the scalar
curvature, and for fermions, which correspond to adding $- \dfrac 16 R$
and $- \dfrac 14 R$ respectively [2], the topological charge
in curved spacetime
is not quantized.  This remains true because these additions will 
not affect the form of inhomogeneity
of (equation 1). Because this equation is inhomogenous, 
we were able to find out that the topological charge for fermions cannot be 
quantized.  Similarly, since the topological charge in curved spacetime 
for fermions and bosons follow the same form, we conclude that 
the charge is not quantized in either case.
Thus, we conclude that the topological charge is quantized for
bosons and not quantized for fermions whereas the instanton solution could not
distinguish between the two cases.  

\noindent
Acknowledgement:
The author would like to thank Professor Alan Chodos of Yale University 
for his assistance and for the useful comments from which this 
manuscript has benefitted.

\bibliographystyle{amsplain}

\end{document}